\documentclass[aps,amsmath,amssymb,prl,twocolumn,showpacs,nofootinbib]{revtex4-1}
\usepackage{bm}
\usepackage{epsfig}
\usepackage[usenames]{color}

\newcommand{\lgl}{\left\langle}
\newcommand{\rgl}{\right\rangle}

\newcommand{\bqu}{\begin{quote}}
\newcommand{\equ}{\end{quote}}

\newcommand{\eqa}{\begin{eqnarray}}
\newcommand{\eqan}{\begin{eqnarray*}}\newcommand{\eqe}{\end{eqnarray}}
\newcommand{\eqen}{\end{eqnarray*}}

\newcommand{\lk}{\left(}    \newcommand{\rk}{\right)}

\newcommand{\lK}{\left[} \newcommand{\rK}{\right]}

\newcommand{\bnabla}{\boldsymbol\nabla}

\newcommand{\cC}{{\mathcal C}}

\newcommand{\cH}{\mathcal  H} 

\newcommand{\bk}{\mathbf k}

\newcommand{\bx}{{\boldsymbol{x}}}

\newcommand{\phibar}{\phi\kern-1ex\rule[1.25ex]{1ex}{.1ex}}

\usepackage{bm}
\usepackage{epsfig}
\usepackage[usenames]{color}
\DeclareFontFamily{OT1}{pzc}{}
\DeclareFontShape{OT1}{pzc}{m}{it}{<-> s * [1.10] pzcmi7t}{}
\DeclareMathAlphabet{\mathpzc}{OT1}{pzc}{m}{it}

\newcommand{\bS}{\mathbf S}

\begin{document}

\title{Vector chiral phases in frustrated 2D XY model and quantum spin chains\footnote{kk}}
\author{H. Schenck$^1$}\author{V.L. Pokrovsky $^{2,3}$}
\author{T. Nattermann$^1$}\affiliation{$^1$Institut f\"ur Theoretische Physik, Universit\"at zu K\"oln, Z\"ulpicher
Str. 77, D-50937 K\"oln, Germany}\affiliation{$^2$Department of Physics, Texas A\&M University, College Station, Texas 77843-4242}\affiliation{$^3$Landau Institute for Theoretical Physics, Chernogolovka, Moscow District, 142432, Russia}

\date{\today}

\begin{abstract}
Abstract: The phase diagram of the frustrated 2D classical and 1D quantum XY models is calculated analytically. 
Four transitions are found: the vortex unbinding transitions  triggered by strong fluctuations occur above and below 
the chiral transition temperature. 
Vortex interaction is short range on small and logarithmic on large scales. 
The chiral
transition,  though belonging to the Ising universality class by symmetry, has a different critical exponents due to non-local interaction.
In a narrow region close to the Lifshitz point a reentrant 
phase transition between paramagnetic and quasi-ferromagnetic phase
appears. 
Applications to antiferromagnetic quantum spin chains and multi-ferroics are discussed.

\end{abstract}

\pacs{75.10.Kt,75.40.Cx,75.50.Ee,75.85.+t}

\maketitle

{\it Introduction.---} 
Landau theory describes phase transitions accompanied by a loss of  symmetry. 
Strong  fluctuations effects   lead either to non-mean-field critical behavior or    first order transition \cite{Chaikin+95}.
The situation is by far less clear in frustrated systems, where discrete and continuous  symmetries can be broken  simultaneously. 
Villain, in a seminal work  \cite{Villain77}, showed that in helical magnets, in addition to  the magnetic order, there exists a second, chiral order parameter  related to the mutual spin orientation on neighboring sites
\eqa\label{chirality}
\boldsymbol{\kappa}=\lgl\bS_i\times\bS_{i+\hat x}\rgl.
\eqe
It soon became  clear that 
many other - interaction or lattice frustrated - models  exhibit this type of order as well (see \cite{Hasenbusch+05,Korshunov06,Balents10} for reviews).

The considerations in the present article are restricted  to  helical magnets for three reasons: 
(i) they are interesting because of their  possible applications as multi-ferroics \cite{Mostovoy06,Cheong+07}, 
(ii) they are sufficiently simple to allow controlled analytical approaches, but 
(iii)  still give a rich phase diagram (see Fig.1).

Villain \cite{Villain78}  considered  a system of $XY$-spins with competing  nearest and  next nearest neighbor interaction along the $\hat x$-axis, which gives rise to helical order with $\boldsymbol \kappa=\pm \kappa \hat x$. \%
Below we use    this helical XY  (HXY) model  as a prototype model of frustrated spin systems. 
It describes  likewise  frustrated  quantum spin chains  at zero temperature, which can be mapped to   1+1-dimensional classical  spin models, provided the spin $S$ is large enough \cite{Kolezhuk02}.

 Using  mean-field analysis Villain \cite{Villain78} found 
a chirally ordered phase { above} the  transition where magnetic order disappears. 
Whereas in three dimensions more sophisticated renormalization group (RG) methods indicate the existence of a single transition \cite{Kawamura98}, the situation is significantly more complicated in two dimensions. 
Here the condensation of topological defects as vortices and  domain walls are expected  to be relevant mechanisms.
 Garel and Doniach \cite{Garel+80}  mapped the HXY-model  to two coupled $XY$-models, resulting in a phase diagram with the chiral transition  below the magnetic transition, in contrast to \cite{Villain78}. 
 However their mapping procedure is doubtful (see \cite{Kolezhuk00}). 
 Okwamoto  \cite{Okwamoto84} used a self-consistent harmonic approximation (SCHA), which yield a phase diagram of the same  topology as in \cite{Garel+80}, but the dependence of the transition lines on the pitch of the helix is  different. 
Kolezhuk  used   simple estimates for the energy of the topological defects in a 1+1-dimensional quantum spin chain to  find an Onsager-like chiral transition above the XY-transition, tacitly assuming that the Ising order parameter has a standard local Hamiltonian \cite{Kolezhuk00}.  
Most part of the analytical work on quantum spin chains is restricted to the $S=1/2$ case, where the mapping to our classical model is questionable, or to parameter regions far from those considered in this work \cite{Nersesyan+98,Lecheminant+01}.

Different numerical approaches have been used as well. Hikihara et al. \cite{Hikihara+00} considered a spin-1  chain using  density matrix RG and  obtained, depending on  frustration, gapped and gapless chiral phases. 
These  correspond in the 2D  classical case to  magnetically disordered and quasi-long range ordered phases, respectively.
In  extended Monte-Carlo studies on the 2D classical HXY model, Cinti et al. \cite{Cinti+11} and Sorokin and Syromyatnikov  \cite{Sorokin+12}  found   transition lines and critical exponents. 
However they used an inappropriate finite size scaling analysis \cite{Ferrenberg+91} which does not take  in account the %
strong anisotropy of systems near the Lifshitz point. As it was shown in \cite{Hornreich+75,Binder+89} such anisotropy requires a strong modification of the scaling analysis.
Nevertheless, it is possible to extract from their raw data 
exponents which turn out to be close to ours (see below).

The new features we have found in this   model, which distinguish our work from all preceding literature, are: 

(i) The non-locality of the chiral order fluctuations leading to strong modifications of its critical behavior in comparison to 2D Ising model possessing the same symmetry. 
(ii) The strong anisotropy of the model leading to different scaling behavior in different direction, as known from Lifshitz points. 
(iii) The anomalous large core of the vortices leads to strong modification of the vortex fugacity.

Our investigation reveals a remarkable simple picture: 
In the  helical ground state both  U(1)  and  parity symmetry are broken.
 With the degeneracy space  SO$(2)\times {\mathbb Z}_2$, relevant excitations are spin waves, vortices and domain walls. 
 Generically, domain walls consist of a regular array of magnetic vortices \cite{Li+12}. 
At low temperatures spin waves  reduce the magnetic order to quasi-long range (algebraic) order.
Spin wave interaction on scales small compared to the chiral correlation length $\xi$ results in non-classical critical exponents  at the chiral transition.  
Vortices on these scales do not interact.
On scales larger than $\xi$ the role of spin waves and vortices interchanges: spin wave interaction becomes irrelevant whereas the    vortex interaction is logarithmic. 
Reduction of the chiral order at increasing temperatures lowers the energy of vortices, resulting in  the Berezinskii-Kosterlitz-Thouless (BKT) transition \cite{Chaikin+95} before the chiral transition takes place. 
Both phases exhibit a non-zero vector chirality  (\ref{chirality}).
Close to the Lifshitz point there appears a reentrant transition to a quasi-ferromagnetic phase (see Fig.1). 
\begin{figure}[htb]
\begin{center}
\includegraphics[width=8.5cm]
{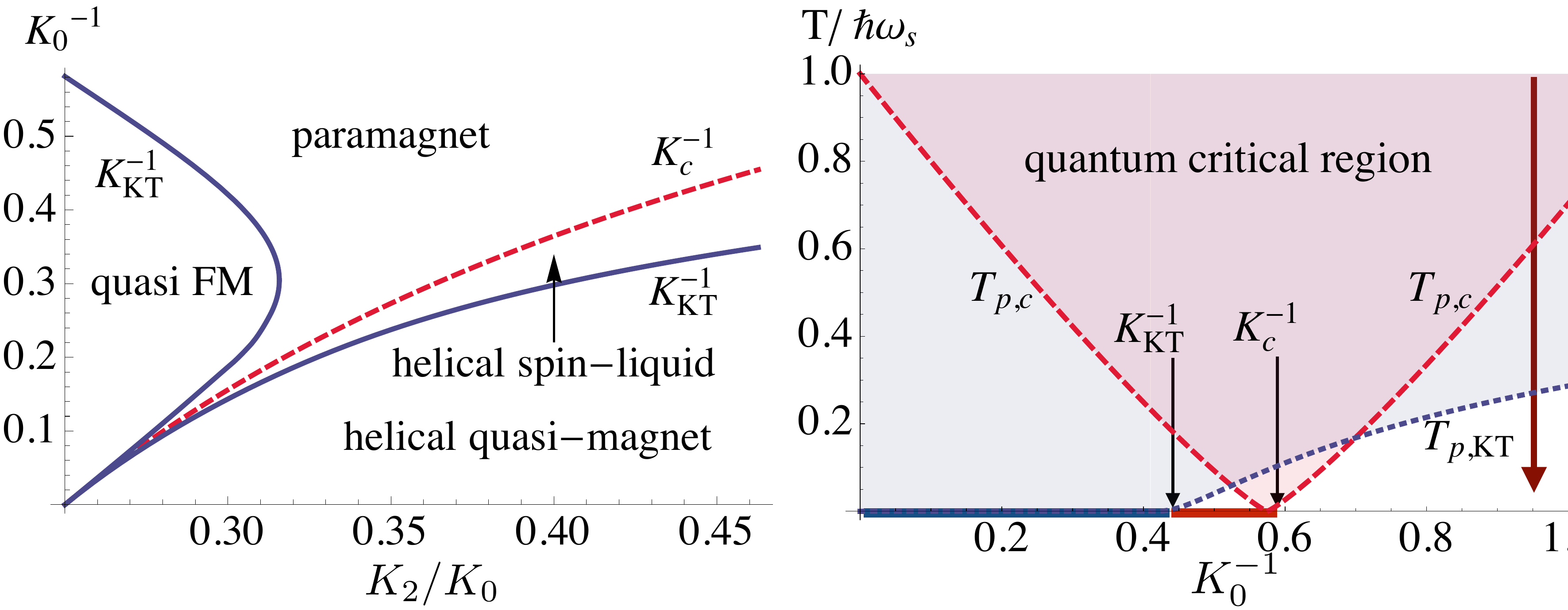}
\caption{Left panel: Phase diagram of the  HXY model as a function of $K_2/K_0$, calculated from (\ref{BKT-transition}). The bold lines mark the BKT-transition, the dashed line the chiral transition. Right panel: Quantum critical regions, $T>T_{p,c}$ for the chiral and $T>T_{p,\textrm{KT}}$ for the magnetic transition, respectively,  of a frustrated quantum spin chain. The thick arrow denotes the parameter region accessible in Gd(hfac)$_3$NITiPr. }
\label{default}
\end{center}
\end{figure}

{\it The model.---}
 In this paper we will consider the { classical} anisotropic XY-model on a square lattice \cite{Villain59} 
\eqa
\label{Hamiltonian_K-K'}
\overline \cH=-\sum\nolimits_{i}\lk K_0\bS_i\bS_{i+\hat x}+K_1\bS_i\bS_{i+\hat y}-K_2\bS_i\bS_{i+2\hat x}\rk.
\eqe 
Here $K_n=J_n/T$, $K_1,K_2>0$,  $\hat x ,\hat y$ denotes the unit vector in $x, y$ direction and  the lattice spacing is set equal to unity. 
In terms of the parameter  $k=K_0/4K_2$, 
the ground state   of (\ref{Hamiltonian_K-K'}) is either  
 ferromagnetic ($1<k$),  {helical magnetic} ($ -1<k<1$) or { anti-ferromagnetic} ($k<-1$). 
At $k=0$ the system decays into two independent sub-lattices which undergo separate BKT-transition.
 Since the Hamiltonian is invariant under the change $K_0\!\to\! -K_0$  and simultaneously flipping all spins on one sublattice of the bi-partite lattice, the results for $K_0<0$ can be obtained from that for $K_0>0$, to which we restrict ourselves now. 
 With $\bS_i=(\cos\phi_i,\sin\phi_i)$ the Hamiltonian can be expressed in terms of $\phi_{i+\hat x}-\phi_i\to\partial_x \phi\equiv\phi_x$ etc. 
Assuming for simplicity $K_1=|K_0|$ we get
\eqa\label{H2}
\overline\cH\!\!=K_0\int_\bx\!\!\!\lK \tfrac{1}{4k}\cos \lk 2\phi_x\rk-\cos\phi_x+\tfrac{1}{8k}\phi^2_{xx}+\tfrac{1}{2}\phi_y^2\rK\!,
\eqe
where $\int_\bx=\int dxdy$. With the Ansatz $\phi_x=\theta$ the energy is minimized by $
\theta=\pm\arccos k.
$ 
Below $\theta$  will be considered  as a small parameter ensuring the validity of continuous approximation. %
Then eq. (\ref{H2}) simplifies to  
  \eqa
  \label{mmHamiltonian-phi1}
\overline\cH=\frac{K_0}{2}\int_\bx\lK -\tfrac{1}{2}\theta^2\phi_x^2 +\phi_y^2+\tfrac{1}{4}\lk\phi_{xx}^2+\phi_x^4\rk\rK.
\eqe

{\it  Perturbation theory.---}
At low temperatures, $ K_0\theta^2\gtrsim 1$, 
 the Hamiltonian  (\ref{mmHamiltonian-phi1}) can be expanded around one of the minima.  
With $\phi=\pm\theta x+\varphi$ we get
\eqa
\label{low-T-hamiltonian}
{\overline\cH}=\frac{K_0}{2}\int_\bx\lK \theta^2\varphi_x^2+\varphi_y^2\pm\theta\varphi_x^3+\tfrac{1}{4}(\varphi_{xx}^2+\varphi_x^4)\rK.
\eqe
For a simple estimate of the anharmonic terms we ignore the compact nature of $\varphi$  and use  $\varphi_x^3\approx 3\varphi_x \sigma^2
$ and $\varphi_x^4\approx 6\varphi_x^2 \sigma^2$ where $\sigma^2=\lgl\varphi_x^2\rgl$. 
Hence
\begin{align}
\label{eq:Hamiltonian-psi2}
\overline\cH\approx\frac{K_0}{2}\int_\bx\lK\lk \theta^2+\tfrac{3}{2}\sigma^2\rk\lk\varphi_x -\delta \theta\rk^2+\varphi_y^2+\tfrac{1}{4}\varphi_{xx}^2\rK.
\end{align}
 $\delta \theta={\mp3\theta\sigma^2}/({2\theta^2+3\sigma^2})$  
represents a temperature dependent  correction to the wave vector $\pm\theta\hat x$,  reducing the modulation. 
The critical coupling constant $K_0=K_c$, at which the chiral symmetry is restored, can be estimated from $\delta\theta= \mp O(\theta)$.
Alternatively one can  start with the chirally symmetric phase.  
To lowest order in the anharmonicity, $-\theta^2$ in (\ref{mmHamiltonian-phi1}) is replaced by $2r_0=-\theta^2+3\sigma^2(r_0)$,  
\eqa
\sigma^2(r)
=\frac{1}{4\pi^2K_0}\int{dk_xdk_y  k_x^2}\lK{rk_x^2+k_y^2+\tfrac{1}{4}k_x^4}\rK^{-1}.
\eqe
With $\sigma(0)=\cC_1 K_0^{-1}$, $\cC_1=0.73$ one gets   
\eqa
{r}_0\approx\theta^2{t},\quad{t}=\tfrac{K_c}{K_0}-1,\quad K_c(\theta)\approx 3\cC_1/ \theta^2.\eqe 
A variational calculation gives equivalent results.

{\it Vortices.---} 
So far we have neglected the compact nature of the $\phi$ field. 
For the further discussion we  replace   (\ref{mmHamiltonian-phi1}) by the effective Hamiltonian 
 \eqa
 \label{eq:Hamiltonian-phi2}
{\overline\cH}=\frac{K_0}{2}\int_\bx\lk r\phi_x^2 +\phi_y^2+\tfrac{1}{4}\phi_{xx}^2+\tfrac{u}{4}N[\phi_x^4]\rk.
\eqe 
 $N[\phi_x^4]=\phi_x^4-6\sigma^2\phi_x^2+\sigma^4$ denotes the normal product. 
Apparently, $\xi_x=r^{-1/2}$ and $\xi_y=r^{-1}$ play the role of the correlation length parallel to the $x$- and $y$-direction, respectively. 
To discuss the nature of vortices, we consider  a region of area $L_xL_y$ containing a single  vortex.
Rescaling  the coordinates according to $x/L_x\to x$, $y/L_y\to y$, 
the linearized saddle point equation 
 reads
 \eqa
 \label{saddlepoint}
\lambda^2_x\phi_{xx}
-\tfrac{1}{4}\lambda_x^4\phi_{xxxx}+\lambda_y^2\phi_{yy}=0,\quad \lambda_\alpha={\xi_\alpha}/{L_\alpha}.  
 \eqe
Here we ignored a term  $\sim u\lk\phi_x^2-\sigma^2\rk\phi_{xx}$ since the effective (unrescaled) coupling  $u_{\textrm{eff}}\sim  L^{-1/2}$   vanishes on large scales  (see below). 

$\phi$ can  then be decomposed into a spin wave and a vortex contribution,
$\phi=\phi^{(sw)}+\phi^{(v)}, $ which do not interact. 
$\phi^{(sw)}$ carries the chiral order and
 will be  treated in the RG calculation  below. 
 Since in a vortex configuration $\phi^{(v)}$ is of the order unity,  all derivatives in (\ref{saddlepoint}) are also of order unity. 
The vortex solution of eq. (10) is different in two limiting range of length scales:

(i)  On small scales,  $\lambda_\alpha\gg 1$, 
 one can ignore the first term on the lhs of (\ref{saddlepoint}).
  A variational calculation of the vortex configuration with the ansatz  
  \eqa
  \phi(x,y)=f(\zeta)\theta(x)+\lK\pi-f(\zeta)\rK\theta(-x), 
  \eqe
  where $  f(\zeta)=\arcsin \zeta$ and $\zeta={ y}/{\sqrt{\varkappa^2x^4+y^2}}$,
  gives $\varkappa\!=\!0.42$ and the vortex energy $\overline{E}_{\textrm{core}}\!=\!2.38 K_0$.
 From the fact that the energy of these  vortices 
 is  dominated by {\it small} scales we conclude that the interaction between the vortices is {\it short}-range, in contrast to the BKT scenario.
In this case  screening of vortices by vortex pairs on smaller scales is { absent}.
  The vortex density  is of the order $ e^{- \overline{E}_{\textrm{core}}}$ and hence of the order $e^{-5.24/\theta^2}\ll 1$ below the chiral transition.

   (ii) In the opposite case of large scales, $\lambda_\alpha\ll1$,  the second term on the lhs of (\ref{saddlepoint}) is negligible. 
With the choice $\lambda_x=\lambda_y$ we get  standard  BKT vortices as solutions.

{\it RG calculation.---}
The separation of length scales used in the previous paragraph is also relevant to the RG analysis. 
On small scales, $\xi_\alpha\gg L_\alpha$,  
 spin waves strongly interact, implying    non-classical critical exponents at the chiral transition. 
On the contrary, on large scales, $\xi_\alpha\ll L_\alpha$, spin wave interaction becomes irrelevant whereas  vortex interaction leads to the BKT scenario. 

We begin  with the  scales  $\xi_\alpha\gg L_\alpha$. $\phi_x\equiv\psi$ plays the role of the order parameter.  
(\ref{eq:Hamiltonian-phi2}) has  the form of a soft spin $\psi^4$ Ising model, apart from the second term in (\ref{eq:Hamiltonian-phi2}) which can be written as a non-local gradient term 
\eqa
\label{non-local-term}
-\frac{1}{2}\int dxdx'dy|x-x'|\psi_y(x,y)\psi_y(x',y).
\eqe
Therefore the critical exponents are expected to  be in  universality class  different from  the  Onsager exponents. 

We use the standard derivation of the RG flow equations \cite{Chaikin+95} for $r$,  $u$, $K$ and the dimensionless vortex fugacity $z$.
Their initial values, defined on the scale of the lattice constant, are $r_0 \ll 1,\,u_0=1$, $K_0$ 
and $z_0=\exp({-\overline{E}_{\textrm{core}}})$. 
  To make the model amendable to an $\epsilon=(5/2)-d$ expansion we replace $y$ by a $(d-1)$-dimensional vector $\bf y$. 
We first integrate out fluctuations $\phi_\bk$  of wave vectors limited by  inequalities 
$ \pi^2> (k_x^4/4)+k_y^2>\pi^2 e^{-2\ell}
$ 
 and then rescale according to  
$
x=x'e^{\ell/2},\, y=y'e^\ell,\,
r'=re^\ell $. 
$\phi$ as a compact variable as well as $u$ are { not} rescaled.
This leads to the flow equations
 \eqa
\label{RG}
\frac{d\ln u}{d\ell}=- \frac{\cC_2u}{K_0},\quad
\frac{d\ln r}{d\ell}=1-\frac{\cC_2u}{3{K_0}}.
 \eqe 
here $\cC_2=9/(2\pi^3)$. Since there is no vortex interaction on these scales,  $K$ and $z$  changes only  due to rescaling, i.e. ${d\ln K}/{d\ell}=-\epsilon
,\,\,{d\ln z}/{d\ell}={3}/{2}$. %
The rescaling of $z=\exp\lk-\overline{E}_{\textrm{core}}+S\rk$ follows from the vortex entropy  $S=\ln[xy/(x'y')]=3\ell/2$. 
The RG stops at $r_{\ell_c}\approx 1$ where $e^{\ell_c}\equiv\xi_y$. 
Integration of (\ref{RG}) between $\ell=0$ and $\ell=\ell_c$ gives   for $\xi_y$ 
\eqa
\label{r(ell)}
\xi_y=\frac{2}{{t}\theta^2}{\cal T}^{1/3},
\quad
{\cal T}(\xi_y)=1+\frac{\cC_2 }{\epsilon K_0}\lk \xi_y^{\epsilon}-1\rk.
\eqe  
For $({\cC_2}/{\epsilon K_0}) \xi_y^{\epsilon}\gg 1$, i.e. inside the critical region of the chiral transition, one finds  $\xi_y\sim \theta^{-2}|{t}_c/{t}|^{\nu_y}$.  To order $\epsilon$, 
\eqa
\nu_y^{-1}=(2\nu_x)^{-1}=\gamma^{-1}=1-{\epsilon}/{3}.
\eqe
$\nu_\alpha$ denotes the correlation length exponent in the $\alpha$-direction. 
With $K_0\approx K_c$ we obtain in two dimensions 
$
\cC_2/({\epsilon K_c})
\approx0.13\, \theta^2 $ and hence
$
{t}_c\approx 
0.034\,\theta^2$ for the size of the critical region.  
 The  specific heat exponent  $\alpha=\nu_y\epsilon/3$  obeys the  hyper-scaling relation \cite{Hornreich+75} 
 \eqa
 \label{hyperscaling}
 \nu_x+(d-1)\nu_y=2-\alpha  
 \eqe
   which applies to the anisotropic system considered here. 
We have also calculated the exponents $\eta_{x,y}$  defined by the critical propagator ${\cal G}^{-1}(\bk)=(k_x^{4-\eta_x}/4)+k_y^{2-\eta_y}$
 and found to order $\epsilon^2$ $\eta_x=-0.212\epsilon^2$ and $\eta_y=0$.
As expected, all exponents are { different} from the Onsager values $\alpha=0,\nu=1, \eta=1/4$.

On larger scales,  $\ell>\ell_c$, the non-linear term in (\ref{eq:Hamiltonian-phi2}) is irrelevant. 
Since $r(\ell_c)=1$, 
the effective model on this scale is the standard XY-model. 
The RG flow equations in two dimensions are those of BKT \cite{Chaikin+95},
\eqa
\label{BKT-equation1}
{d}K_0^{-1}/{d\ell}=
4\pi^3{z^2},\,\quad
{d\ln z}/{d\ell}=2-\pi K_0.
\eqe
where we use isotropic rescaling. These equations have to be integrated with the initial conditions on the scale $e^{\ell_c}$ 
Thus  $ K_{\ell_c}=K\exp({-\epsilon\ell_c})$ and 
 $z_{\ell_c}\approx \exp({3\ell_c/2-\overline E_{\textrm{core}}})$.  
Integration of  the flow equations   
gives  
 the following relation for the BKT transition temperature 
\eqa\label{BKT-transition}
{2}/({\pi K_{\ell_c}})=1+\ln{2}+2\pi^2z^2_{\ell_c}-\ln({\pi K_{\ell_c}}).
\eqe

{\it The gapped chiral phase.---}
Below $T_c=J/K_c$ we rewrite $\phi_x=\kappa+\varphi_x$ where $\kappa=\lgl\phi_x\rgl\ll 1$.
 The   expansion of the free energy  density with respect to $\kappa$ can be written as
\eqa
\label{free-energy}
\overline{\cal F}=\tfrac{1}{2}{K_0}\lK r_0{\cal T}^{-1/3}\kappa^2+\tfrac{1}{4} {\cal T}^{-1}\kappa^4\rK.
\eqe
Minimization of $\cal F$ gives 
$
\kappa^2={-2r_0}{\cal T}^{2/3}.
$ 
The  correlation length is therefore given again by (\ref{r(ell)}) provided ${t}$ is replaced by $2|{t}|$.
This gives 
\eqa
\kappa\equiv \lgl\phi_x\rgl\sim |{t}|^\beta, \qquad \beta=(1-\epsilon)\nu_x.
\eqe
Our exponents fulfil the scaling relation $\alpha+2\beta+\gamma=2$.

As already mentioned, previous numerical analysis    has ignored the strong anisotropy of the system \cite{Cinti+11,Sorokin+12} which changes the finite size scaling analysis \cite{Binder+89}. 
By the procedure used in \cite{Sorokin+12} most probably  the larger of the two correlation length exponents  is obtained, i.e. $\nu_y =2\nu_x\approx 1$. Then, according to (\ref{hyperscaling}),  $\alpha\approx 1/2$, whereas using the standard scaling relation with no anisotropy,  $\alpha \approx 0.115$ was found in \cite{Sorokin+12}.
However, a direct examination of the temperature plots for the specific heat and the order parameter (Figs.7, 19 of \cite{Sorokin+12}) gives 
$\alpha\approx 0.32,\,\beta\approx0.30$, suggesting $\gamma\approx 1.08$, in reasonable agreement with our values $\alpha=1/6, \beta=1/3, \gamma=7/6$ when expanded to first order in $\epsilon=1/2.$ 
%

{\it Phase-diagram.---}
At low temperatures we have long range chiral order and a power law decay of spin correlations.
This is the chiral nematic (gapless) phase considered in \cite{Nersesyan+98,Kolezhuk00}. 
 Increasing  $K_0^{-1}$  
the numerical solution of (\ref{BKT-transition})  shows that there is a BKT transition  below the chiral transition (see Fig.1), in qualitative agreement with numerical results \cite{Sorokin+12}. 
It is important to note that for finding the correct phase boundary the contribution of small scale free vortices ($\ell<\ell_c$) is essential.
For $\theta\ll 1$, $K_c^{-1},K_{\textrm{ KT}}^{-1}\sim\theta^2$, in agreement with \cite{Okwamoto84} (but the opposite sequence of transitions was found there).

Above the BKT transition the spin correlations are short range,  
the correlation length $\xi_{\textrm{\scriptsize KT}}\approx e^{ 1.5/\sqrt{t_\textrm{KT}}}$ is of the order of the vortex distance.  Here ${t}_\textrm{KT}={K_\textrm{KT}}/{K_0}-1.$
The chiral order parameter vanishes at $T_c=JK_c^{-1}$ that is  slightly larger than $T_{\textrm{\scriptsize KT}}=JK^{-1}_{\textrm{\scriptsize KT}}$.

In the region $0.25\!<\!K_2/K_0\!<\!0.316$ there is a { reentrant} phase transition to the quasi-ferromagnetic phase (see Fig.1).  
It should  however be taken into account that our approach is restricted to small $\theta$. Thus the size of the reentrant region may be overestimated when going to larger $\theta$ values.
Reentrant behavior  was seen before  using SCHA \cite{Okwamoto84}. 
However, the SCHA cannot consider vortices accurately and ignores completely the vortex structure on scales smaller $\xi$.  
%

{\it Antiferromagnetic quantum spin chains}.---
Using the standard mapping, antiferromagnetic ($J_0<0$) spin-$S$ chains at {\it zero} temperature are described by   1+1 classical systems (\ref{mmHamiltonian-phi1}) with  the replacements   
\eqa
K_0= \sqrt{{2}/{3}}\,S, \,\,\ y= v_s\tau,\,\, v_s={\omega_S a },
\eqe
 provided $S\gg1$ \cite{Kolezhuk02}.  
 $\tau$ denotes the imaginary time, $a=1$ the lattice constant,   $v_s $ the spin-wave velocity,  and $\omega_S={\sqrt{{3}/{2}}|\theta J_0|S}/\hbar$.
For increasing $K_0$ the  spin chain undergoes two quantum phase transitions (QPT): at $K_c$    from a paramagnetic to helical spin liquid and at $K_{\textrm{\scriptsize KT}}$  to a quasi-long range ordered magnetic phase (see Fig.1). %
The dynamical critical exponents at the QPT  follows from the relation $\xi_y\sim\xi_x^z$ as	   $z=2-\eta_x/2$ at the chiral, and $z=1$ at the BKT transition, respectively \cite{Sondhi+97}.

Adding a weak interchain coupling $ J_\perp=\varepsilon_\perp  |J|$,  $ \varepsilon_\perp\ll1$,  the system is  equivalent to a higher-dimensional classical system. The latter     presumably undergoes  a single  phase transition \cite{Kawamura98} to  a long range ordered magnetic phase. The transition happens at    ${t}={t}_{3D}$ which  follows from the condition \cite{Scalapino+75}
$
1\approx 2 J_\perp \chi_\textrm{magn}
$ where $\chi_\textrm{magn}$ denotes the magnetic susceptibility of the 1D chain.  
This gives
${t}_{3D}\sim 1/\ln^2\varepsilon_\perp$, in agreement with a more elaborate  RG calculation  \cite{Hikami+80}.

\smallskip

At  low but {\it  finite} $T$,  the imaginary time $\tau$ is restricted to the region $0<\tau<\hbar/T$, i.e. $y<L_y=\hbar \omega_S/T$. 
At $\varepsilon_\perp=0$ the system is now equivalent to a one-dimensional classical model and hence no true phases transition can occur. %
Finite size scaling   gives for the susceptibility \cite{Sondhi+97}
  \eqa\label{finite-size-scaling}
  \chi(K_S,L_y)=\frac{S^2}{|J_0|}\lk\frac{\hbar\omega_S}{T}\rk^{2-\eta_y}\tilde\chi \lk\frac{ \hbar\omega_S}{T\xi_y}\rk. 
  \eqe
In the quantum critical domain, where $\hbar\omega_S\lesssim T\xi_y$ and  $\tilde\chi(x)\approx\tilde\chi(0)$,   $\chi\sim T^{-2+\eta_y}$.

 At the chiral transition,  with $\chi_\textrm{chiral}\sim\int_\bx\lgl\psi(\bx)\psi(0)\rgl$, one finds from $\xi_y\approx {|t|}^{-\nu_y}$ and $\eta_y=0$  that $\chi_\textrm{chiral}$ grows as $\sim T^{-2}$  before reaching  a maximum at   $T\approx T_{p,c}\sim\hbar\omega_S{|t|}^{\nu_y}.$ 
 At the BKT transition where $2-\eta_y=7/4$  
one obtains analogously $\chi_{\textrm{magn}}\sim T^{-7/4}$ at  
$T\gtrsim T_{p,\textrm{\scriptsize{KT}}}\sim\hbar\omega_S \exp(- 1.5/\sqrt{t_\textrm{KT}})$.

For non-zero interchain coupling, the transition temperature for the magnetic transition is found from   $1=2J_\perp\chi_{1D}$ and   (\ref{finite-size-scaling})  as  
\eqa
\label{T3D}
T_{3D}\approx \hbar\omega_S\lK \varepsilon_\perp S^2\tilde\chi\lk e^{- 1.5/\sqrt{t_\textrm{KT}}} \hbar\omega_S/T_{3D}\rk \rK^{\frac{1}{2-\eta}}.
\eqe
At the BKT transition of the chains, where ${t}_{KT}=0$,   $T_{3D}\sim  \hbar\omega_S\varepsilon_\perp^{4/7}$
 which is smaller than the peak temperature $T_p$ by a factor 
 $\varepsilon_\perp^{4/7}\ll 1$.  Our result for the $\hbar\omega_S$ and $\varepsilon_\perp$ dependence of  $T_{3D}$ agrees with that found in \cite{Affronte+99} if the mean field exponent $2-\eta_y=2$ is used. Other details differ since in \cite{Affronte+99} spin wave theory was used    within the chains.

 \smallskip
{\it Experiments}.---
There is a large number of rare earth metals, alloys and compounds which exhibit helical phases \cite{Jensen+91,Kimura+08,Chapon+04}. Unfortunately experiments on films to our knowledge where done  only for cases where the helical axis is perpendicular to the film plane \cite{Weschke+04}. The other group of materials to which our theory applies are frustrated  quantum spin with large  S.   
In Gd(hfac)$_3$NITiPr half of the spins are S=7/2
and hence sufficiently large, as required, the other half are $S=1/2$ such that $S_\textrm{eff}\approx \sqrt{7}/2$.  
Two peaks at $T_N=1.88$ K and $T_c=2.19$ K were indeed found in the specific heat of this material \cite{Cinti+08}, which were interpreted as the magnetic and chiral transition, respectively.  In contrast our theory explains these peaks as quantum critical phenomena.
With $J_0\approx 7.06$ K, $\varepsilon_\perp\approx 2.3\times10^{-3}$,  and $\theta\approx 0.36 \pi$ \cite{Cinti+08} one finds $\hbar\omega_S\approx 12.87$ K, $K_0=1. 08$,  $K_c=1.74$, and $K_{\textrm{\scriptsize{KT}}}\approx 2.44$,  i.e. at zero temperature the single chains are  in their paramagnetic phase (see Fig.1).  $T_{3D}\approx  0.55$K is much smaller than the observed peak temperatures. The latter are given only up to prefactors of order unity as    $T_{p,c}\approx 7.23\, $ K and $T_{p,\textrm{\scriptsize{KT}}}\approx 3.39\, $K. The values of the prefactors follow  from a  calculation of $\tilde\chi(x)$, which is beyond the scope of this article.

The other chain compounds have spin S=1/2, resulting in a competition of dimerization and frustration. In some of them the effect of the frustration is dominating. An example is LiCu$_2$O$_2$ where two nearby transitions have been found as well \cite{Matsuda+04}.

In multiferroics the electric polarisation $\bf P$   is coupled to the magnetisation according to ${\bf P}\sim ({\mathbf m}\cdot\bnabla){\mathbf m}-{\mathbf m}(\bnabla\cdot{\mathbf m})\sim\gamma\hat x$ \cite{Mostovoy06}.
 Long range chiral order should be therefore detectable by measuring $\bf P$.
 
 \smallskip
 
\noindent{ The authors thank O. Dimitrowa   for  interesting  discussions.
This work has been supported by  University of Cologne Center of Excellence QM2 and by the DOE under
the grant DE-FG02-06ER 46278.}%

\bibliography{chiral-1}
\end{document}